\documentclass[twocolumn]{aastex63}

\newcommand\teff{\mbox{$T_\mathrm{eff}$}}
\usepackage{color}

\begin{document}

\title{Substellar Hyades Candidates from the UKIRT Hemisphere Survey}

\correspondingauthor{Adam C. Schneider}
\email{aschneid10@gmail.com}

\author[0000-0002-6294-5937]{Adam C. Schneider}
\affil{United States Naval Observatory, Flagstaff Station, 10391 West Naval Observatory Rd., Flagstaff, AZ 86005, USA}
\affil{Department of Physics and Astronomy, George Mason University, MS3F3, 4400 University Drive, Fairfax, VA 22030, USA}

\author{Frederick J. Vrba}
\affil{United States Naval Observatory, Flagstaff Station, 10391 West Naval Observatory Rd., Flagstaff, AZ 86005, USA}

\author[0000-0002-4603-4834]{Jeffrey A. Munn}
\affil{United States Naval Observatory, Flagstaff Station, 10391 West Naval Observatory Rd., Flagstaff, AZ 86005, USA}

\author[0000-0002-2968-2418]{Scott E. Dahm}
\affil{United States Naval Observatory, Flagstaff Station, 10391 West Naval Observatory Rd., Flagstaff, AZ 86005, USA}

\author[0000-0002-3858-1205]{Justice Bruursema}
\affil{United States Naval Observatory, Flagstaff Station, 10391 West Naval Observatory Rd., Flagstaff, AZ 86005, USA}

\author[0000-0002-3858-1205]{Stephen J. Williams}
\affil{United States Naval Observatory, Flagstaff Station, 10391 West Naval Observatory Rd., Flagstaff, AZ 86005, USA}

\author[0000-0002-5604-5254]{Bryan N. Dorland}
\affil{United States Naval Observatory, 3450 Massachusetts Ave NW, Washington, DC 20392-5420, USA}

\author[0000-0001-6251-0573]{Jacqueline K. Faherty}
\affil{Department of Astrophysics, American Museum of Natural History, Central Park West at 79th St., New York, NY 10024, USA}

\author[0000-0003-4083-9962]{Austin Rothermich}
\affil{Department of Astrophysics, American Museum of Natural History, Central Park West at 79th St., New York, NY 10024, USA}

\author{Emily Calamari}
\affil{Department of Astrophysics, American Museum of Natural History, Central Park West at 79th St., New York, NY 10024, USA}

\author[0000-0001-7780-3352]{Michael C. Cushing}
\affil{Ritter Astrophysical Research Center, Department of Physics and Astronomy, University of Toledo, 2801 W. Bancroft St., Toledo, OH 43606, USA}

\author[0000-0001-7896-5791]{Dan Caselden}
\affil{Department of Astrophysics, American Museum of Natural History, Central Park West at 79th St., New York, NY 10024, USA}
\author[0000-0003-4905-1370]{Martin Kabatnik}
\affil{Backyard Worlds: Planet 9, USA}
\author{William Pendrill}
\affil{Backyard Worlds: Planet 9, USA}
\author[0000-0003-4864-5484]{Arttu Sainio}
\affil{Backyard Worlds: Planet 9, USA}
\author[0000-0003-4714-3829]{Nikolaj Stevnbak Andersen}
\affil{Backyard Worlds: Planet 9, USA}
\author{Christopher Tanner}
\affil{Backyard Worlds: Planet 9, USA}

\begin{abstract}

We have used data from the UKIRT Hemisphere Survey (UHS) to search for substellar members of the Hyades cluster.  Our search recovered several known substellar Hyades members, and two known brown dwarfs that we suggest may be members based on a new kinematic analysis.  We uncovered thirteen new substellar Hyades candidates, and obtained near-infrared follow-up spectroscopy of each with IRTF/SpeX.  Six candidates with spectral types between M7 and L0 are ruled out as potential members based on their photometric distances ($\gtrsim$100 pc).  The remaining seven candidates, with spectral types between L5 and T4, are all potential Hyades members, with five showing strong membership probabilities based on BANYAN $\Sigma$ and a convergent point analysis.   Distances and radial velocities are still needed to confirm Hyades membership. If confirmed, these would be some of the lowest mass free-floating members of the Hyades yet known, with masses as low as $\sim$30 $M_{\rm Jup}$.  An analysis of all known substellar Hyades candidates shows evidence that the full extent of the Hyades has yet to be probed for low-mass members, and more would likely be recovered with deeper photometric and astrometric investigations.     

\end{abstract}


\section{Introduction}
\label{sec:intro}

Brown dwarfs have central temperatures that never reach the critical threshold for stable thermonuclear H burning \citep{kumar1963, hayashi1963}.  These substellar objects thus do not form a main sequence, but instead radiatively cool over time, thereby following a mass--luminosity--age relationship. It is therefore difficult to constrain brown dwarf fundamental properties such as mass, luminosity, or age, because one of them must be known to determine the other two. Brown dwarfs with known ages, while rare, can break this degeneracy.  For this reason, any brown dwarf that can be tied to a nearby young association or open cluster with a well-constrained age provides a valuable benchmark for fundamental tests of substellar theory.  

The Hyades is the closest open cluster to the Sun ($\sim$47 pc; \citealt{lodieu2019}).  As such, it has been extensively characterized, resulting in well-determined member identification down to the substellar boundary (e.g., \citealt{roeser2011, gaia2018, reino2018, lodieu2019, smart2021}),  a well-determined age of $\sim$650 Myr (e.g., \citealt{lebreton2001, degennaro2009, martin2018, lodieu2019}), and an established slightly supersolar metallicity (e.g., [Fe/H]$\sim$0.146 dex; \citealt{cummings2017}). As the nearest open cluster, Hyades members also have significant proper motions ($\mu_{\rm total}$ $\sim$100 mas yr$^{-1}$).  While some properties of the Hyades cluster, such as its distance and its relatively large proper motion, make it an ideal site for investigations of substellar populations, there are limitations to its full exploration.  Being so near, Hyades members extend over a very large area of the sky, with tidal tails extending even further \citep{meingast2019, roeser2019}, making surveys with deep imaging of the entire cluster challenging.  While large scale infrared surveys have enabled some exploration of the nearer cluster members (e.g., \citealt{perez2017}), such surveys (e.g., 2MASS; \citealt{skrutskie2006}) are not deep enough to detect substellar members with very low temperatures over the entire Hyades distance range.  Despite these challenges, several L and T type members and candidate members of the Hyades have been identified \citep{bouvier2008, hogan2008, perez2017, schneider2017, perez2018, zhang2021}. 

We have performed a large area search for new candidate substellar members of the Hyades using the United Kingdom Infra-Red Telescope (UKIRT) Hemisphere Survey (UHS; \citealt{dye2018}), which covers the majority of the spatial extent of the Hyades.  We describe our search in Section \ref{sec:targets} and follow-up spectroscopic observations in Section \ref{sec:obs}.  The analysis of our new Hyades substellar candidates is presented in Section \ref{sec:anal}, and a discussion of our results is given in Section \ref{sec:disc}.

\section{Target Selection}
\label{sec:targets}

The UKIRT Hemisphere Survey (UHS) covers approximately 12,700 deg$^2$ in the northern hemisphere.  Combined with existing UKIDSS surveys \citep{lawrence2007}, the UHS covers the entire northern hemisphere between 0$\degr$ and 60$\degr$.  The $J-$band portion of the survey has been publicly released \citep{dye2018}, and the $K-$band survey has an anticipated public release some time in 2023.  

We constructed a proper motion catalog based on UHS data by cross-matching each $K-$band UHS detection with the UHS $J-$band catalog, after first removing those sources from each UHS catalog with matches in {\it Gaia} EDR3 \citep{gaia2021}.  The $J-$/$K-$band matching was done with incrementally increasing matching radii.  For each match, a preliminary proper motion was calculated by differencing the $J-$ and $K-$band positions.  Matches were kept only if the $K-$band detection had a corresponding entry in CatWISE 2020 \citep{marocco2021}, after propagating the $K-$band position to the CatWISE 2020 epoch using the preliminary proper motion.  Final proper motions were calculated based on the $J-$ and $K-$band positions, as well as the Pan-STARRS (PS1) DR2 \citep{chambers2016, magnier2020} position for those objects with a match in the Pan-STARRS catalog.

Recent studies using {\it Gaia} data have led to the discovery of Hyades tidal tails \citep{meingast2019, roeser2019}.  The identification of members of the Hyades tidal tails necessitated a spatial density filter, which requires accurate distances.  Since we expect any new candidates found through our search to be beyond {\it Gaia} magnitude limits, we focus our search for substellar Hyades candidates around the cluster center and omit the recently discovered Hyades tidal tails.  To select candidates from our UHS proper motion catalog, we limit our search to objects within 18 pc of the cluster center, which should include all bound and halo cluster members \citep{lodieu2019}.  To do this, we imposed right ascension and declination constraints based on the extremes of known halo Hyades members from the {\it Gaia} Catalogue of Nearby Stars (GCNS; \citealt{smart2021}).  Note that the \cite{smart2021} census includes 560 of the 568 Hyades members within 18 pc of the cluster center found in \cite{lodieu2019}.  Specifically, we required 46 $\leq$ R.A.~(deg) $\leq$ 85 and $-$4.5 $\leq$ Dec.~(deg) $\leq$ 38.5. We also ensured each candidate had proper motion components consistent with known Hyades members from \cite{lodieu2019} by imposing proper motion constraints based on these same members: 42 $\leq$ $\mu_{\alpha}$ (mas yr$^{-1}$) $\leq$ 197 and $-$92 $\leq$ $\mu_{\delta}$ (mas yr$^{-1}$) $\leq$ 43.  To identify substellar candidates, we select only sources with $J-$W2 colors $>$ 1.5 mag, which is inclusive of the vast majority of L and T type brown dwarfs (see e.g., Figure 7 of \citealt{kirkpatrick2016}).  We also chose a $J-$band magnitude limit of 17.5 mag, which corresponds to the approximate limit of what is observable with the SpeX spectrograph \citep{rayner2003} in prism mode at NASA's Infrared Telescope Facility (see Section \ref{sec:obs}).  Over 450 candidates were selected via these criteria.  

We verified each source was a kinematic match to the Hyades using the BANYAN~$\Sigma$ classifier \citep{gagne2018}, keeping those with a non-zero probability of membership in the Hyades.  BANYAN~$\Sigma$ uses sky positions, proper motions, and, when available, radial velocities and distances to determine the probability that a given object is a member of any nearby young association or cluster using Bayesian statistics.  There were 105 objects that returned a non-zero BANYAN~$\Sigma$ probability of belonging to Hyades.  A visual inspection of each object further reduced the number of candidates to 25, where candidates removed via visual inspection were typically blended or extended objects.  

\begin{deluxetable*}{lcrrcccccc}
\label{tab:rec}
\tablecaption{Recovered Substellar Hyades Members}
\tablehead{
\colhead{CWISE} & \colhead{Disc.} & \colhead{$\mu_{\alpha}$} & \colhead{$\mu_{\delta}$} & \colhead{$J_{\rm UHS}$\tablenotemark{a}} & \colhead{$K_{\rm UHS}$\tablenotemark{a}} & \colhead{SpT} & \colhead{SpT} \\
\colhead{Name} & \colhead{Ref.} & \colhead{(mas yr$^{-1}$)} & \colhead{(mas yr$^{-1}$)} & \colhead{(mag)} & \colhead{(mag)} & \colhead{} & \colhead{Ref.} }
\startdata
J035304.34$+$041820.0\tablenotemark{b} & 1 & 171.2$\pm$3.1 & 35.8$\pm$2.9 & 16.312$\pm$0.013 & 14.509$\pm$0.010 & L6pec (red) & 1\\
J041232.79$+$104408.0\tablenotemark{c} & 2 & 129.5$\pm$3.8 & -5.5$\pm$3.5 & 17.471$\pm$0.041 & 15.263$\pm$0.019 & L5:~(red) & 2\\
J041733.97$+$143015.2 & 3 & 123.6$\pm$2.7 & -17.8$\pm$2.3 & 16.468$\pm$0.015 & 14.625$\pm$0.011 & L2\tablenotemark{d} & 4\\
J041835.00$+$213126.6 & 5 & 142.0$\pm$4.3 & -51.8$\pm$4.0 & 17.203$\pm$0.027 & 15.195$\pm$0.015 & L5 & 5\\
J042418.72$+$063745.5 & 6 & 138.8$\pm$2.8 & 7.5$\pm$2.9 & 17.222$\pm$0.022 & 15.434$\pm$0.019 & L4 & 6\\
J043038.87$+$130956.7 & 7 & 141.6$\pm$3.5 & -21.5$\pm$3.6 & 16.869$\pm$0.017 & 16.199$\pm$0.037 & T2.5 & 8\\
J043543.04$+$132344.8 & 3 & 95.7$\pm$3.9 & -17.7$\pm$3.4 & 16.714$\pm$0.016 & 14.892$\pm$0.012 & L6 (red)\tablenotemark{e} & 9\\
J043642.79$+$190134.6 & 2 & 113.5$\pm$2.0 & -42.1$\pm$2.0 & 16.766$\pm$0.017 & 14.849$\pm$0.013 & L6 & 2\\
J043803.58$+$070055.2 & 6 & 88.7$\pm$3.1 & 3.1$\pm$3.1 & 16.778$\pm$0.018 & 14.993$\pm$0.014 & L1 & 6\\
J043855.29$+$042300.6 & 10 & 118.7$\pm$3.5 & 11.7$\pm$3.4 & 16.381$\pm$0.012 & 15.108$\pm$0.016 & T2 & 10\\
J044105.60$+$213001.3 & 2 & 98.7$\pm$4.6 & -48.5$\pm$4.4 & 17.441$\pm$0.032 & 15.352$\pm$0.021 & L5 (red) & 2\\
J044635.44$+$145125.7 & 3 & 79.1$\pm$2.6 & -22.4$\pm$2.5 & 16.378$\pm$0.015 & 14.594$\pm$0.009 & L3.5 & 11 \\
\enddata
\tablenotetext{a}{Based on the UKIRT photometric system \citep{hodgkin2009}}
\tablenotetext{b}{CWISE J035304.34$+$041820.0 has not been suggested as a potential Hyades member before this work.}
\tablenotetext{c}{CWISE J041232.79$+$104408.0 was suggested to be an unlikely Hyades member in \cite{schneider2017}, though we find it to be high-probability member using more precise astrometry.}
\tablenotetext{d}{\cite{lodieu2014} report an optical spectral type of L1 for this object, while \cite{martin2018} found an optical spectral type of L3.5.  \cite{lodieu2019} gives a spectral type of L2, and we adopt that type here.}
\tablenotetext{e}{\cite{lodieu2014} found an optical spectral type of L3.5 for this object.}
\tablerefs{ (1) \cite{kellogg2017}; (2) \cite{schneider2017}; (3) \cite{hogan2008}; (4) \cite{lodieu2019}; (5) \cite{perez2017}; (6) \cite{perez2018}; (7) \cite{bouvier2008}; (8) \cite{liu2016}; (9) \cite{best2015}; (10) \cite{best2020}; (11) \cite{martin2018}}
\end{deluxetable*}

\subsection{Recovered Substellar Hyades Candidates}
\label{sec:recovered}

Of the remaining 25 candidates, 11 were previously suggested substellar Hyades members from \cite{bouvier2008}, \cite{hogan2008}, \cite{perez2017}, \cite{schneider2017}, \cite{perez2018}, and \cite{zhang2021}.  Details of these recovered members are listed in Table \ref{tab:rec}.  

One object recovered by our search was CWISE J041232.79$+$104408.0 (WISEA J041232.77$+$104408.3; \citealt{schneider2017}).  CWISE J041232.79$+$104408.0 was discovered in \cite{schneider2017} with a spectral type of L5:~(red), and was considered an unlikely Hyades member based on a crude proper motion and a comparison with known Hyades members at that time.  Our measured proper motion components for this source have smaller uncertainties than the \cite{schneider2017} values by a factor of $\sim$10 (see Table \ref{tab:rec}).  We therefore reevaluate this object's potential membership based on these updated values and BANYAN $\Sigma$ and find a 96.2\% probability of Hyades membership.  

We further test the potential Hyades membership of CWISE J041232.79$+$104408.0 by evaluating whether or not its proper motion is consistent with the Hyades convergent point, defined in \cite{madsen2002}.  Following \cite{hogan2008}, we compare the proper motion angle ($\theta_{\mu}$) and the angle measured between a line pointing north and a line from CWISE J041232.79$+$104408.0 to the convergent point ($\theta_{\rm cp}$), which should be similar for Hyades members.  We find $\theta_{\mu}$ = 92\fdg4 and $\theta_{\rm cp}$ = 93\fdg6.   Considering our proper motion precisions, these angles are discrepant by $<$1$\sigma$.  We can also use the measured proper motion of CWISE J041232.79$+$104408.0 and the moving cluster method to estimate a distance to this object if it were a Hyades member.  Using the cluster velocity of 46.38 km s$^{-1}$ from \cite{lodieu2019}, we find a distance of 42.2 pc, which agrees well with this object's photometric distance estimate of 46$\pm$6 pc from \cite{schneider2017}. We thus reclassify CWISE J041232.79$+$104408.0 as a likely Hyades member.

Another object recovered by our search was CWISE J043803.58$+$070055.2 (2M0438$+$0700), which was suggested as a Hyades candidate in \cite{perez2018}, with a spectral type of L1.  Using their spectral type and our $K-$band magnitude for this source, we find a photometric distance of $\sim$77 pc, which is much larger than the $\sim$44 pc distance estimate from \cite{perez2018}.  If CWISE J043803.58$+$070055.2 is $\sim$77 pc distant, it is $>$30 pc from the Hyades cluster center.  One possible explanation that would bring CWISE J043803.58$+$070055.2 closer to the cluster center is a later spectral type, as the spectrum used to type this object in \cite{perez2018} covers a limited wavelength range and has a low signal-to-noise ratio (S/N).  A high S/N spectrum of this source may be warranted.  We include it in Table \ref{tab:rec} as a possible Hyades member for completeness.

\begin{figure*}
\plotone{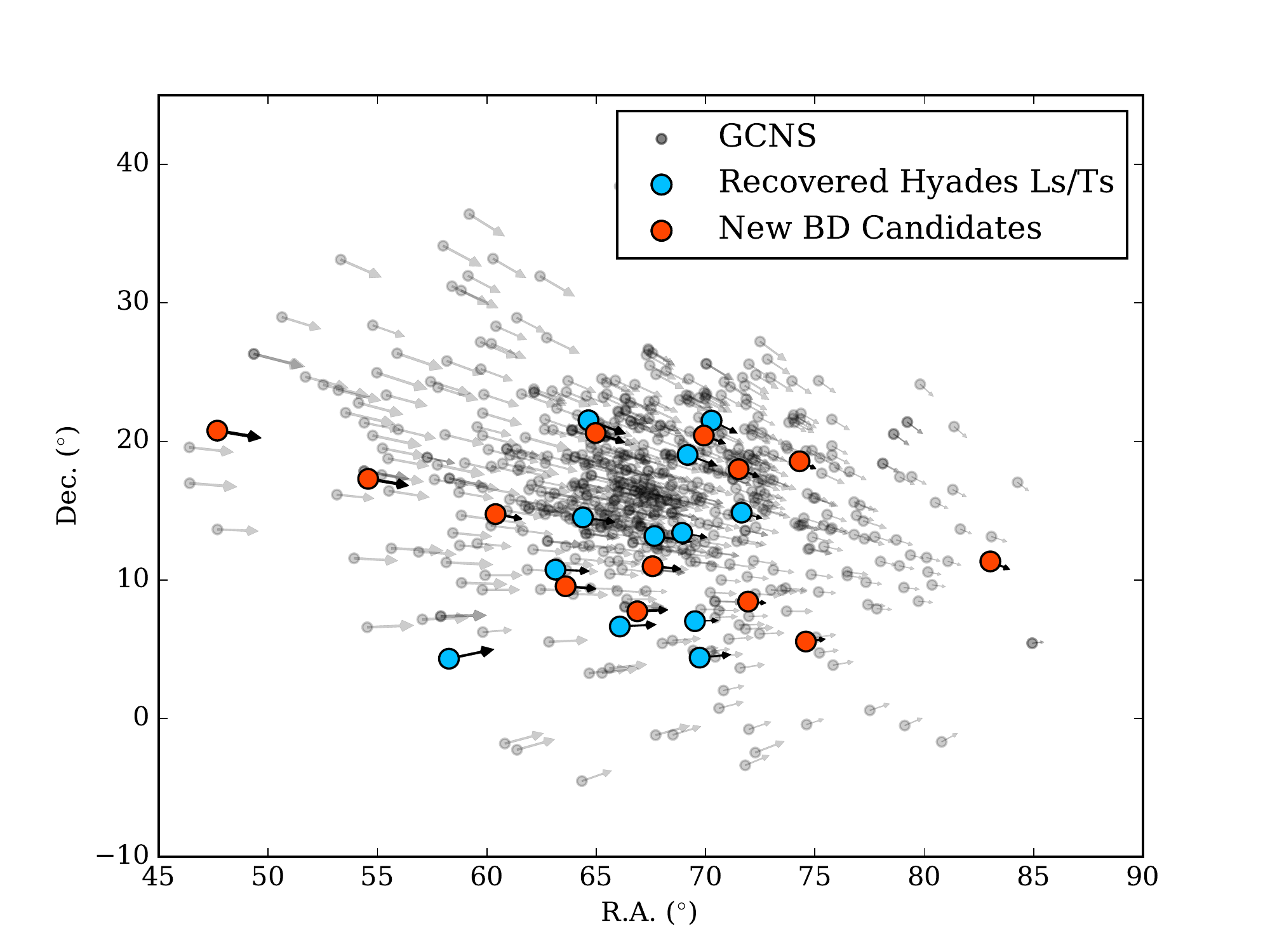}
\caption{The J2000 positions and proper motion vectors of our substellar Hyades candidates (red) and known, recovered L and T type members of the Hyades (blue) compared to all known Hyades members within the halo radius (18 pc) from the GCNS \citep{smart2021}.}  
\label{fig:plot1}
\end{figure*}

\subsection{Other Recovered Brown Dwarfs}
\label{sec:other}

We also recovered the known brown dwarf CWISE J035304.34$+$041820.0 (2MASS J03530419$+$0418193; \citealt{kellogg2017}). CWISE J035304.34$+$041820.0 is an extremely red L dwarf discovered in \cite{kellogg2017} which has not been linked to the Hyades previously.  Many known, young brown dwarfs appear redder than field age counterparts with similar spectral types (e.g., \citealt{faherty2016}). And while CWISE J035304.34$+$041820.0 is extremely red compared to other objects with an L6 spectral type, \cite{kellogg2017} noted that this source did not display any spectroscopic signatures of youth.  The low-gravity features used to diagnose young ages for brown dwarfs are calibrated for ages $\lesssim$200 Myr (e.g., \citealt{allers2013}), and therefore the lack of youthful features in the spectrum of CWISE J035304.34$+$041820.0 does not rule out Hyades membership. 

Using our measured proper motion for this source from Table \ref{tab:rec}, we find a 14.7\% chance of belonging to the Hyades cluster from BANYAN $\Sigma$ and a kinematic distance assuming Hyades membership of $\sim$34 pc.  \cite{schneider2016} found that $K-$band photometric distances show good agreement with measured parallaxes for exceptionally red objects, especially compared to other photometric bands.  Using the absolute $K-$band magnitude versus spectral type relation from \cite{dupuy2012} and the UHS $K-$band magnitude for this source, we find a photometric distance of $\sim$31 pc, in good agreement with the kinematic distance estimate.  At a distance of $\sim$31 pc, CWISE J035304.34$+$041820.0 would be $\sim$19 pc from the Hyades cluster center given in \citep{lodieu2019}, just beyond the halo region defined in that work (18 pc).  

\begin{longrotatetable} 
\begin{deluxetable*}{lrrccccccccccc}
\label{tab:props}
\tablecaption{New Substellar Hyades Candidates}
\tablehead{
\colhead{CWISE} & \colhead{$\mu_{\alpha}$} & \colhead{$\mu_{\delta}$} & \colhead{$i_{\rm PS1}$} & \colhead{$z_{\rm PS1}$} & \colhead{$y_{\rm PS1}$} & \colhead{$J_{\rm UHS}$\tablenotemark{a}} & \colhead{$K_{\rm UHS}$\tablenotemark{a}} & \colhead{W1} & \colhead{W2}  \\
\colhead{Name} & \colhead{(mas yr$^{-1}$)} & \colhead{(mas yr$^{-1}$)} & \colhead{(mag)} & \colhead{(mag)} & \colhead{(mag)} & \colhead{(mag)} & \colhead{(mag)} & \colhead{(mag)} & \colhead{(mag)}}
\startdata
J031042.59$+$204629.3\tablenotemark{b} & 164.7$\pm$2.2 & -27.7$\pm$2.2 & 20.702$\pm$0.031 & 19.240$\pm$0.013 & 18.228$\pm$0.013 & 15.956$\pm$0.011 & 14.267$\pm$0.009 & 13.723$\pm$0.014 & 13.435$\pm$0.013  \\
J033817.87$+$171744.1\tablenotemark{c} & 152.0$\pm$3.5 & -25.3$\pm$3.3 & \dots & 20.789$\pm$0.075 & 19.707$\pm$0.034 & 17.311$\pm$0.026 & 15.185$\pm$0.015 & 14.315$\pm$0.016 & 13.910$\pm$0.015   \\
J040136.03$+$144454.6 &  99.3$\pm$3.3 & -18.2$\pm$2.6 & 20.860$\pm$0.048 & 19.512$\pm$0.027 & 18.604$\pm$0.022 & 16.840$\pm$0.025 & 15.706$\pm$0.028 & 15.512$\pm$0.023 & 15.314$\pm$0.041 \\
J041424.22$+$093223.5 & 114.5$\pm$3.6 & -9.8$\pm$3.5 & 21.052$\pm$0.067 & 19.814$\pm$0.029 & 18.948$\pm$0.038 & 17.274$\pm$0.035 & 16.278$\pm$0.045 & 15.945$\pm$0.028 & 15.654$\pm$0.053  \\
J041953.55$+$203628.0\tablenotemark{d} & 109.4$\pm$9.0 & -35.8$\pm$8.9 & \dots & \dots & 19.827$\pm$0.081 & 17.290$\pm$0.029 & 16.990$\pm$0.078 & 16.243$\pm$0.032 & 15.390$\pm$0.040  \\
J042731.38$+$074344.9\tablenotemark{e} & 114.3$\pm$3.5 &  5.5$\pm$3.1 & \dots & 20.555$\pm$0.031 & 19.496$\pm$0.066 & 17.188$\pm$0.025 & 15.302$\pm$0.019 & 14.242$\pm$0.016 & 13.875$\pm$0.015  \\
J043018.70$+$105857.1 & 106.3$\pm$6.9 & -10.7$\pm$6.9 & \dots & 21.034$\pm$0.182 & 19.903$\pm$0.143 & 17.451$\pm$0.027 & 17.313$\pm$0.115 & 16.855$\pm$0.050 & 15.828$\pm$0.062 \\
J043941.41$+$202514.8 & 80.8$\pm$8.0 & -30.3$\pm$7.9 & \dots & 21.594$\pm$0.185 & 19.844$\pm$0.123 & 17.453$\pm$0.027 & 17.196$\pm$0.101 & 16.451$\pm$0.038 & 15.464$\pm$0.043  \\
J044603.23$+$175930.8 & 77.3$\pm$5.1 & -28.8$\pm$4.7 & 21.440$\pm$0.050 & 19.993$\pm$0.042 & 19.205$\pm$0.033 & 17.453$\pm$0.024 & 16.418$\pm$0.049 & 15.897$\pm$0.027 & 15.571$\pm$0.050  \\
J044747.32$+$082552.0 & 66.2$\pm$3.5 & -5.3$\pm$3.6 & 21.090$\pm$0.058 & 19.772$\pm$0.024 & 18.854$\pm$0.023 & 16.956$\pm$0.024 & 15.782$\pm$0.030 & 15.433$\pm$0.022 & 15.144$\pm$0.033  \\
J045712.03$+$183344.1 & 61.0$\pm$2.5 & -28.0$\pm$2.6 & 21.256$\pm$0.040 & 19.847$\pm$0.046 & 19.002$\pm$0.028 & 17.254$\pm$0.025 & 16.132$\pm$0.037 & 15.910$\pm$0.029 & 15.751$\pm$0.054  \\
J045821.05$+$053244.5 & 72.1$\pm$2.8 & 9.4$\pm$2.7 & 20.831$\pm$0.027 & 19.476$\pm$0.024 & 18.457$\pm$0.019 & 16.600$\pm$0.016 & 15.440$\pm$0.022 & 15.125$\pm$0.020 & 14.929$\pm$0.027  \\
J053204.60$+$111955.1\tablenotemark{f} & 72.4$\pm$1.9 & -30.2$\pm$1.9 & 21.257$\pm$0.053 & 19.420$\pm$0.033 & 18.366$\pm$0.032 & 15.944$\pm$0.009 & 13.932$\pm$0.006 & 13.193$\pm$0.015 & 12.764$\pm$0.011  \\
\enddata
\tablenotetext{a}{Based on the UKIRT photometric system \citep{hodgkin2009}}
\tablenotetext{b}{CWISE J031042.59$+$204629.3 was independently discovered by citizen scientist Nikolaj Stevnbak.}
\tablenotetext{c}{CWISE J033817.87$+$171744.1 was independently discovered by citizen scientists Christopher Tanner, Sam Goodman, and Martin Kabatnik.}  
\tablenotetext{d}{CWISE J041953.55$+$203628.0 was independently discovered by citizen scientist Martin Kabatnik.}
\tablenotetext{e}{CWISE J042731.38$+$074344.9 was independently discovered by citizen scientists Dan Caselden and Billy Pendrill.}
\tablenotetext{f}{CWISE J053204.60$+$111955.1 was independently discovered by citizen scientists Arttu Sainio and Sam Goodman.}
\end{deluxetable*}
\end{longrotatetable} 

As with CWISE J041232.79$+$104408.0, we compared the proper motion angle of CWISE J035304.34$+$041820.0 to the convergent point angle and found $\theta_{\mu}$ = 78\fdg2 and $\theta_{\rm cp}$ = 84\fdg4.  Considering our proper motion precisions, these angles are discrepant by $\sim$3$\sigma$.  Using the moving cluster method, we find a kinematic distance of 35.2 pc.  This distance matches reasonably well with our photometric distance estimate of 31 pc.  We therefore consider 2MASS J03530419$+$0418193 a potential Hyades member.   A parallax and radial velocity for this source would help to firmly establish Hyades membership.  

\subsection{New Substellar Hyades Candidates}
The remaining 13 candidates are listed in Table \ref{tab:props}, which includes photometry from UHS \citep{dye2018}, CatWISE 2020 \citep{marocco2021} and Pan-STARRS (PS1) DR2 \citep{chambers2016, magnier2020}.  We show the positions of these 13 candidates compared to known cluster and halo Hyades members from \cite{smart2021} and our recovered substellar members from Table \ref{tab:rec} in Figure \ref{fig:plot1}.  We note here that 5 of these thirteen candidates were independently discovered by citizen scientists working with the Backyard Worlds: Planet 9 project \citep{kuchner2017}.  These citizen scientists are recognized in the table notes of Table \ref{tab:props}.

\section{Observations}
\label{sec:obs}

\subsection{IRTF/SpeX}
\begin{figure*}
\plotone{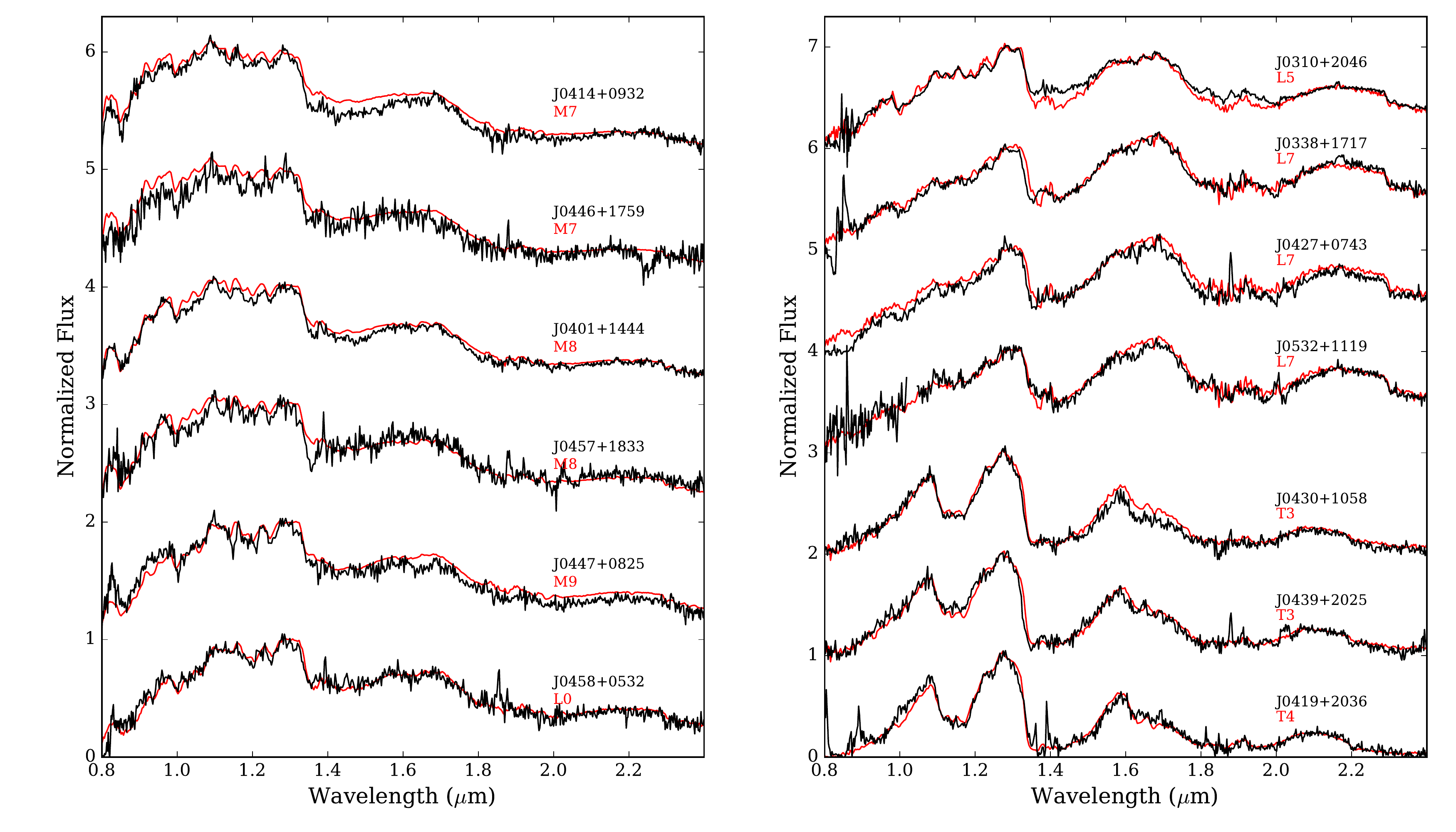}
\caption{IRTF/SpeX spectra (black) compared to spectral standards (red).  The spectra are normalized between 1.27 and 1.29 $\mu$m and offset by integer values for clarity.  The spectral standards are: VB 8 (M7; \citealt{burgasser2008}); VB 10 (M8; \citealt{burgasser2004}); LHS 2924 (M9; \citealt{burgasser2006b}); 2MASP J0345432$+$254023 (L0; \citealt{burgasser2006b}); SDSS J083506.16$+$195304.4 (L5; \citealt{chiu2006});  2MASSI J0103320$+$193536 (L7; \citealt{cruz2004}); 2MASS J12095613$-$1004008 (T3; \citealt{burgasser2004}); 2MASSI J2254188$+$312349 (T4; \citealt{burgasser2004}).}  
\label{fig:spectra}
\end{figure*}

We obtained near-infrared spectra of our 13 substellar Hyades candidates with the SpeX spectrograph \citep{rayner2003} at NASA's 3 m Infrared Telescope Facility (IRTF) on UT 2021 Nov 11 and 12.  The observations were taken in prism mode with the 0\farcs8 slit, which gives a spectral resolution of $\lambda/\Delta\lambda$ $\approx$150 across the 0.8$-$2.4 $\mu$m wavelength range.  A0 stars were observed immediately after each target for telluric correction purposes.  Because these observations were concentrated on the Hyades, some A0 star observations were suitable for multiple targets.   Depending on the brightness of the target, we took between 2 and 16 images of 180 seconds each in an ABBA pattern with the slit aligned to the parallactic angle.  Calibration files were taken between the target and the telluric observations, and the spectral extraction, wavelength calibration, and telluric correction were performed with the SpeXTool package \citep{vacca2003, cushing2004}.   Details of the observations can be found in Table \ref{tab:obs}, and the final, reduced spectra are shown in Figure \ref{fig:spectra}.  We also give the S/N at the $J-$band peak for our reduced spectra in Table \ref{tab:obs}.

\begin{deluxetable*}{lcccccc}
\label{tab:obs}
\tablecaption{IRTF Observations}
\tablehead{
\colhead{CWISE} & \colhead{Obs.\ Date} & \colhead{Total Exp.\ Time} & \colhead{A0 Star} & \colhead{Spec.} & \colhead{(S/N)$_J$} \\
\colhead{Name} & \colhead{(UT)} & \colhead{(s)} & \colhead{} & \colhead{Type} & \colhead{}  }
\startdata
J031042.59$+$204629.3 & 2021 Nov 11 & 1440 & HD19600 & L5 & 81 \\
J033817.87$+$171744.1 & 2021 Nov 11 & 2160 & HD 35036 & L7 & 45 \\
J040136.03$+$144454.6 & 2021 Nov 11 & 2160 & HD 35036 & M8 & 62 \\
J041424.22$+$093223.5 & 2021 Nov 11 & 2880 & HD 35036 & M7 & 39 \\
J041953.55$+$203628.0 & 2021 Nov 11 & 2880 & HD 35036 & T4 & 29 \\
J042731.38$+$074344.9 & 2021 Nov 12 & 900 & HD 31411 & L7 & 33 \\
J043018.70$+$105857.1 & 2021 Nov 12 & 1260 & HD 35036 & T3 & 34 \\
J043941.41$+$202514.8 & 2021 Nov 12 & 1080 & HD 35036 & T3 & 31 \\
J044603.23$+$175930.8 & 2021 Nov 12 & 1080 & HD 35036 & M7 & 19 \\
J044747.32$+$082552.0 & 2021 Nov 12 & 1080 & HD 35036 & M9 & 38 \\
J045712.03$+$183344.1 & 2021 Nov 12 & 1080 & HD 35036 & M8 & 26 \\
J045821.05$+$053244.5 & 2021 Nov 11 & 1800 & HD 31411 & L0 & 27 \\
J053204.60$+$111955.1 & 2021 Nov 11 & 360 & HD 31411 & L7 & 27 \\
\enddata
\end{deluxetable*}

\section{Analysis}
\label{sec:anal}

\subsection{Spectral Types}
\label{sec:spts}

Spectral types for each candidate were determined by comparing $J-$band morphologies to near-infrared spectral standards from \cite{burgasser2006} and \cite{kirkpatrick2010}.  We use standard $\chi^2$ fitting to determine the best fitting standards, and confirm each by-eye.  The best fitting spectral types are given in Table \ref{tab:obs} and the best matching near-infrared spectral standards are shown in Figure \ref{fig:spectra}.  As seen in the table, all candidates have spectral types of M7 or later, with the latest spectral types being T3 and T4.  

\subsection{Distances}
\label{sec:astrometry}

None of our Hyades candidates have detections or distance measurements from {\it Gaia} \citep{gaia2021} or any other astrometric study.  We use the absolute-magnitude vs.\ spectral type relations from \cite{dupuy2012} and our UHS $J-$ and $K-$band photometry to find distance ranges for each of our Hyades candidates, including a spectral type subclass uncertainty of $\pm$0.5.  Distance ranges for each candidate are given in Table \ref{tab:member}.

\subsection{Hyades Membership}

We evaluate our sample for Hyades membership using several methods.  We first use the BANYAN~$\Sigma$ software \citep{gagne2018} using only our measured proper motions from Table \ref{tab:props} and the positions of each object.  We then use the BANYAN~$\Sigma$ classifier with our photometric distance ranges as an input parameter.  BANYAN~$\Sigma$ probabilities are given in Table \ref{tab:member}.  For the six objects that have spectral types earlier than L0 (CWISE J040136.03$+$144454.6, CWISE J041424.22$+$093223.5, CWISE J044603.23$+$175930.8, CWISE J044747.32$+$082552.0, CWISE J045712.03$+$183344.1, and CWISE J053204.60$+$111955.1), their BANYAN~$\Sigma$ Hyades membership probabilities drop to 0\% when their photometric distances are included.  This is not surprising given that their photometric distance estimates are all $\gtrsim$100 pc, well beyond the furthest known Hyades cluster members.  We consider all of these objects non-members, and discuss them further in Section \ref{sec:lateMs}. 

Both CWISE J031042.59$+$204629.3 and CWISE J053204.60$+$111955.1 have relatively small Hyades membership probabilities from BANYAN~$\Sigma$ ($<$50\%), with or without the inclusion of their photometric distance estimates.  We consider both objects possible Hyades members based on their BANYAN~$\Sigma$ probabilities.

For the remaining five objects, their already high ($>$60\%) Hyades membership probabilities from BANYAN~$\Sigma$ increased when their photometric distance estimates were included in their evaluation.  We consider all of these objects strong Hyades candidate members based on their BANYAN~$\Sigma$ analysis.

We also evaluate each candidate's potential Hyades membership using their measured astrometry compared to the Hyades convergent point.  As with CWISE J035304.34$+$041820.0 and CWISE J041232.79$+$104408.0 in Section \ref{sec:other}, we calculate each candidate's proper motion angle ($\theta_{\mu}$) and convergent point angle ($\theta_{\rm cp}$).  These are provided in Table \ref{tab:member}.  All angles are consistent to within $\pm$3$\degr$ for each object, the one exception being CWISE J053204.60$+$111955.1, which has a proper motion angle of 112\fdg6 and a convergent point angle of 106\fdg2.  Our typical proper motion uncertainty is $\pm$4 mas yr$^{-1}$, which corresponds to a proper motion angle uncertainty of $\pm$2\fdg3, making a range of $\pm$7$\degr$ equivalent to $\pm$3$\sigma$.  The angle difference for CWISE J053204.60$+$111955.1 is just within this range, and it is thus not ruled out as a candidate.  

We also include the convergent point distance, and BANYAN~$\Sigma$ predicted distances in Table \ref{tab:member}.  For the 7 objects with BANYAN~$\Sigma$ probabilities greater than 0\% when photometric distances are included, the photometric distances, convergent point distances, and BANYAN~$\Sigma$ predicted distances are reasonably consistent. For CWISE J053204.60$+$111955.1, if it is a Hyades member, it is well outside the halo radius of the cluster.  Using a photometric distance estimate of $\sim$20 pc, CWISE J053204.60$+$111955.1 is $\sim$28 pc from the $xyz$ position of the cluster center given in \cite{lodieu2019}.  \cite{lodieu2019} and \cite{smart2021} identified over 100 Hyades cluster members between 18 and 30 pc from the cluster center, so Hyades membership for CWISE J053204.60$+$111955.1 cannot be ruled out.  Regardless, CWISE J053204.60$+$111955.1 has a photometric distance estimate suggesting that it may be part of the 20 pc sample of nearby stars and brown dwarfs.  

\begin{deluxetable*}{lcrccrrrrc}
\label{tab:member}
\tablecaption{Hyades Membership Summary}
\tablehead{
\colhead{CWISE} & \colhead{Spec.} & \colhead{dist$_{\rm phot}$} & \colhead{dist$_{\rm cp}$} & \colhead{dist$_{\rm BANYAN}$} & \colhead{$\theta_{\mu}$} & \colhead{$\theta_{\rm cp}$} & \colhead{BANYAN\tablenotemark{a}}  & \colhead{BANYAN\tablenotemark{a}}& \colhead{Member?} \\
\colhead{Name} & \colhead{Type} & \colhead{(pc)} & \colhead{(pc)} & \colhead{(pc)} & \colhead{($\degr$)} & \colhead{($\degr$)} & \colhead{(\%)} & \colhead{(\%)} & }
\startdata
J031042.59$+$204629.3 & L5 & 29--36 & 45.2 & 45.2 & 99.6 & 98.7 & 47.7 & 29.6 & Y?\\
J033817.87$+$171744.1& L7 & 33--46 & 43.6 & 43.1 & 99.5 & 98.6 & 83.6 & 89.3 & Y \\
J040136.03$+$144454.6 & M8 & 134--157 & 58.9 & 56.4 & 100.4 & 98.2 & 28.1& 0.0 & N \\
J041424.22$+$093223.5 & M7 & 185--253 & 47.4 & 47.1 & 94.9 & 91.9 & 64.2 & 0.0 & N \\
J041953.55$+$203628.0& T4 & 34--41 & 48.2 & 46.9 & 108.1 & 109.0 & 93.3 & 96.7 & Y \\
J042731.38$+$074344.9 & L7 & 35--43 & 43.4 & 42.3 & 87.3 & 89.6 & 90.0 & 97.7 & Y \\
J043018.70$+$105857.1 & T3 & 41--53 & 45.7 & 45.3 & 95.8 & 95.4 & 94.4 & 98.2 & Y \\
J043941.41$+$202514.8 & T3 & 41--51 & 56.8 & 51.8 & 110.6 & 113.1 & 62.0 & 91.9 & Y \\
J044603.23$+$175930.8 & M7 & 201--274 & 55.2 & 53.2 & 110.4 & 110.5 & 81.9 & 0.0 & N \\
J044747.32$+$082552.0 & M9 & 125--143 & 63.3 & 57.2 & 94.6 & 91.7 & 23.9 & 0.0 & N \\
J045712.03$+$183344.1 & M8 & 160--190 & 62.8 & 59.4 & 114.7 & 114.5 & 27.6 & 0.0 & N \\
J045821.05$+$053244.5 & L0 & 98--109 & 52.3 & 49.1 & 82.6 & 85.4 & 31.5 & 0.0 & N \\
J053204.60$+$111955.1 & L7 & 19--24 & 32.2 & 31.2 & 112.6 & 106.2 & 7.1 & 34.2 & Y?\\
\enddata
\tablenotetext{a}{The first BANYAN Hyades membership probability listed does not include a distance estimate as a constraint, while the second uses the photometric distance to calculate the probability of Hyades membership.}
\end{deluxetable*}

\subsection{Physical Properties}
\label{sec:phys}

Previous studies have shown that field-age and young brown dwarfs with similar spectral types have significantly different effective temperatures (e.g., \citealt{filippazzo2015}).  However, these investigations have generally compared field-age brown dwarfs to brown dwarfs with ages $\lesssim$200 Myr.  \cite{liu2016} showed that the T2.5 Hyades member CFHT-Hy-20 (CWISE J043038.87$+$130956.7) had infrared photometry consistent with the field population.  Thus when estimating effective temperatures for our sample, we use the relation from \cite{kirkpatrick2021} for field age brown dwarfs.    We include a $\pm$0.5 subtype uncertainty for spectral type.  

To estimate masses, we use two sets of models; the hybrid models from \cite{saumon2008} and those from \cite{phillips2020}.  Uncertainties are found in a Monte Carlo fashion, where we assume a normal age distribution around 650$\pm$50 Myr for each object.  Masses and effective temperatures for each potential Hyades member are given in Table \ref{tab:phys}.  The model-estimated masses for each of our 7 possible Hyades members are all below 50 $M_{\rm Jup}$, firmly in the substellar regime.  Our new T-type candidate members all have masses $\sim$30 $M_{\rm Jup}$, which puts them amongst the lowest masses of candidate Hyades members. 

Previously suggested T-type Hyades members include CFHT-Hy-20 and CFHT-Hy-21 \citep{bouvier2008}, with spectral types of T2.5 and T1 respectively \citep{bouvier2008, liu2016} and PSO J049.1159$+$26.8409, PSO J052.2746$+$13.3754, and PSO J069.7303$+$04.3834 \citep{zhang2021}, with spectral types of T2.5, T3.5, and T2, respectively \citep{best2015, best2020}.  \cite{zhang2021} also suggested the known T6.5 brown dwarf WISEPA J030724.57$+$290447.6 \citep{kirkpatrick2011} as a potential Hyades member, but we rule out Hyades membership for this object in Section \ref{sec:census}. Because spectral type scales with mass for substellar objects with the same age, and CWISE J041953.55$+$203628.0 has the latest spectral type of candidate Hyades members (T4), it is likely the lowest mass free-floating Hyades member yet known.

\begin{deluxetable*}{lccccc}
\label{tab:phys}
\tablecaption{Physical Properties on New Hyades Candidates}
\tablehead{
\colhead{CWISE} & \colhead{Spec.} & \colhead{\teff} & \colhead{Mass\tablenotemark{a}} & \colhead{Mass\tablenotemark{b}}  \\
\colhead{Name} & \colhead{Type} & \colhead{(K)} & \colhead{($M_{\rm Jup}$)} & \colhead{($M_{\rm Jup}$)} }
\startdata
J031042.59$+$204629.3 & L5 & 1610$\pm$140 & 55$^{+6}_{-8}$ & 49$\pm$6\\
J033817.87$+$171744.1 & L7 & 1420$\pm$140 & 46$^{+7}_{-12}$ & 41$^{+6}_{-5}$ \\
J041953.55$+$203628.0 & T4 & 1180$\pm$80 & 28$^{+4}_{-2}$ & 32$\pm$3 \\
J042731.38$+$074344.9 & L7 & 1420$\pm$140 & 46$^{+7}_{-12}$ & 41$^{+6}_{-5}$ \\
J043018.70$+$105857.1 & T3 & 1200$\pm$80 & 30$^{+5}_{-3}$ & 33$\pm$3 \\
J043941.41$+$202514.8 & T3 & 1200$\pm$80 & 30$^{+5}_{-3}$ & 33$\pm$3 \\
J053204.60$+$111955.1 & L7 & 1420$\pm$140 & 46$^{+7}_{-12}$ & 41$^{+6}_{-5}$ \\
\enddata
\tablenotetext{a}{Masses determined using the hybrid models of \cite{saumon2008}}
\tablenotetext{b}{Masses determined using the models of \cite{phillips2020}}
\end{deluxetable*}

\section{Discussion}
\label{sec:disc}

\subsection{What about those distant late-Ms/early-Ls with Hyades proper motions?}
\label{sec:lateMs}

Our search returned 6 late-M or early-L type objects with Hyades-like proper motions that have photometric distance measurements well beyond the nominal cluster radius.  While it is intriguing to consider these objects as part of a potential extended Hyades stream, their significant proper motions at their estimated distances suggest very different space velocities than known Hyades members.  Using their photometric distances and proper motions, we find tangential velocity ($V_{\rm tan}$) values ranging from 36 km s$^{-1}$ for CWISE J045821.05$+$053244.5 to 120 km s$^{-1}$ for CWISE J041424.22$+$093223.5, many of which are consistent with the thick disk or halo population of the Milky Way (e.g., \citealt{bensby2003}). For Hyades members within the halo radius of 18 pc from \cite{smart2021}, we find an average $V_{\rm tan}$ of 24$\pm$4 km s$^{-1}$.  We conclude that all of these objects are likely unrelated to the Hyades and are therefore background interlopers.

\subsection{The Current Census of Substellar Hyades Members}
\label{sec:census}

There have been several relatively recent attempts to identify substellar Hyades members (e.g., \citealt{perez2018, zhang2021}).  We took this opportunity to combine the results of these efforts with our search to examine the completeness of these combined investigations.  One might expect a different distance distribution for substellar Hyades members than stellar members, as efforts to identify substellar members are typically magnitude limited in some way.  This is either because of the specific dataset being used to identify candidates, or other considerations, such as the $J <$17.5 mag criterion we imposed in our search to facilitate follow-up observations. To investigate this, we sought to compare the distance distribution of known Hyades members from \cite{smart2021} with known and suspected substellar Hyades members.  For the known Hyades members, we use the list of 713 suggested members from \cite{smart2021} within 30 pc of the Hyades cluster center.  We do not limit to the 18 pc halo radius as we did with our specific UHS brown dwarf search because previous surveys for substellar members have various selection criteria.  Therefore, the full 713 member sample will return a more consistent comparison.  

For the substellar Hyades sample, we include all suspected Hyades members with spectral types of L0 or later, which corresponds to a mass of 72 $M_{\rm Jup}$ at the age of the Hyades using the evolutionary models of \cite{phillips2020}.  We use 11 objects recovered in our search from Table \ref{tab:rec} and our 7 new candidates from Table \ref{tab:member}.  To these samples, we include 4 other L dwarfs from \cite{hogan2008} with spectral types in \cite{lodieu2014} or \cite{martin2018} (Hya02, Hya03, Hya08, and Hya11), one additional T dwarf from \cite{bouvier2008} (CFHT-Hy-21), and two Hyades candidates from \cite{zhang2021} (PSO J049.1159$+$26.8409 and PSO J052.2746$+$13.3754). 

Two previously suggested Hyades members, 2M0429$+$2437 \citep{perez2018} and WISEPA J030724.57$+$290447.6 \citep{zhang2021}, we rule out as potential Hyades members based on a reanalysis of their astrometry.  

2M0429$+$2437 (CWISE J042930.33$+$243749.0) was suggested as a potential Hyades member in \cite{perez2018}, who gave $\mu_{\alpha}$=$+$45 mas yr$^{-1}$ and $\mu_{\delta}$=$-$71 mas yr$^{-1}$ and indicated a typical proper motion uncertainty of $\pm$19.3 mas yr$^{-1}$ for each component.  They also showed a low-S/N spectrum of this source and gave a spectral type of L6--L8, noting that this source was difficult to classify.  CatWISE 2020 \citep{marocco2021} gives $\mu_{\alpha}$=$+$17.2$\pm$14.6 mas yr$^{-1}$ and $\mu_{\delta}$=$-$0.9$\pm$15.3 mas yr$^{-1}$, which are more precise and significantly different than the values given in \cite{perez2018}.  Using the CatWISE 2020 proper motion values of this source, we find a BANYAN~$\Sigma$ Hyades membership probability of 0\% for this source.  An inspection of optical images of the area around this object shows that its colors are likely influenced by a foreground molecular cloud.  We suggest that this object is likely a highly-reddened background object.  Its nature may be illuminated with a higher-S/N spectrum.  

WISEPA J030724.57$+$290447.6 (CWISE J030724.57$+$290447.2) was suggested as a very low mass Hyades member in \cite{zhang2021}.  However, the proper motion components used for this object in that work had significant uncertainties ($\pm$100 mas yr$^{-1}$).  Using the proper motion for this object from our UHS proper motion catalog ($\mu_{\alpha}$=$-$29.8$\pm$14.7 mas yr$^{-1}$ and $\mu_{\delta}$=$-$53.4$\pm$14.6 mas yr$^{-1}$) we find a 0\% BANYAN~$\Sigma$ Hyades membership probability.  

We also exclude CWISE J043803.58$+$070055.2 \citep{perez2018} from our census of potential substellar Hyades members because it is $\gtrsim$30 pc from the Hyades cluster center (see Section \ref{sec:recovered}).

The remaining candidates all have BANYAN~$\Sigma$ membership probabilities for the Hyades $>$0\% (Table \ref{tab:all}).  For this evaluation, we use proper motions with the smallest uncertainties from \cite{liu2016}, \cite{best2018}, \cite{lodieu2019}, {\it Gaia} EDR3 \citep{gaia2021}, or our UHS proper motion catalog.  We also use parallactic distances when available, which come from \cite{liu2016}, \cite{lodieu2019}, \cite{best2020}, and {\it Gaia} EDR3 \citep{gaia2021}.  When a parallax is not available, we use the $K-$band photometric distance.  The general properties of known substellar Hyades candidates are summarized in Table \ref{tab:all}.  We also include a mass estimate for each object following the method outlined in Section \ref{sec:phys} using the evolutionary models of \cite{phillips2020} assuming Hyades membership.   

\begin{figure*}
\plotone{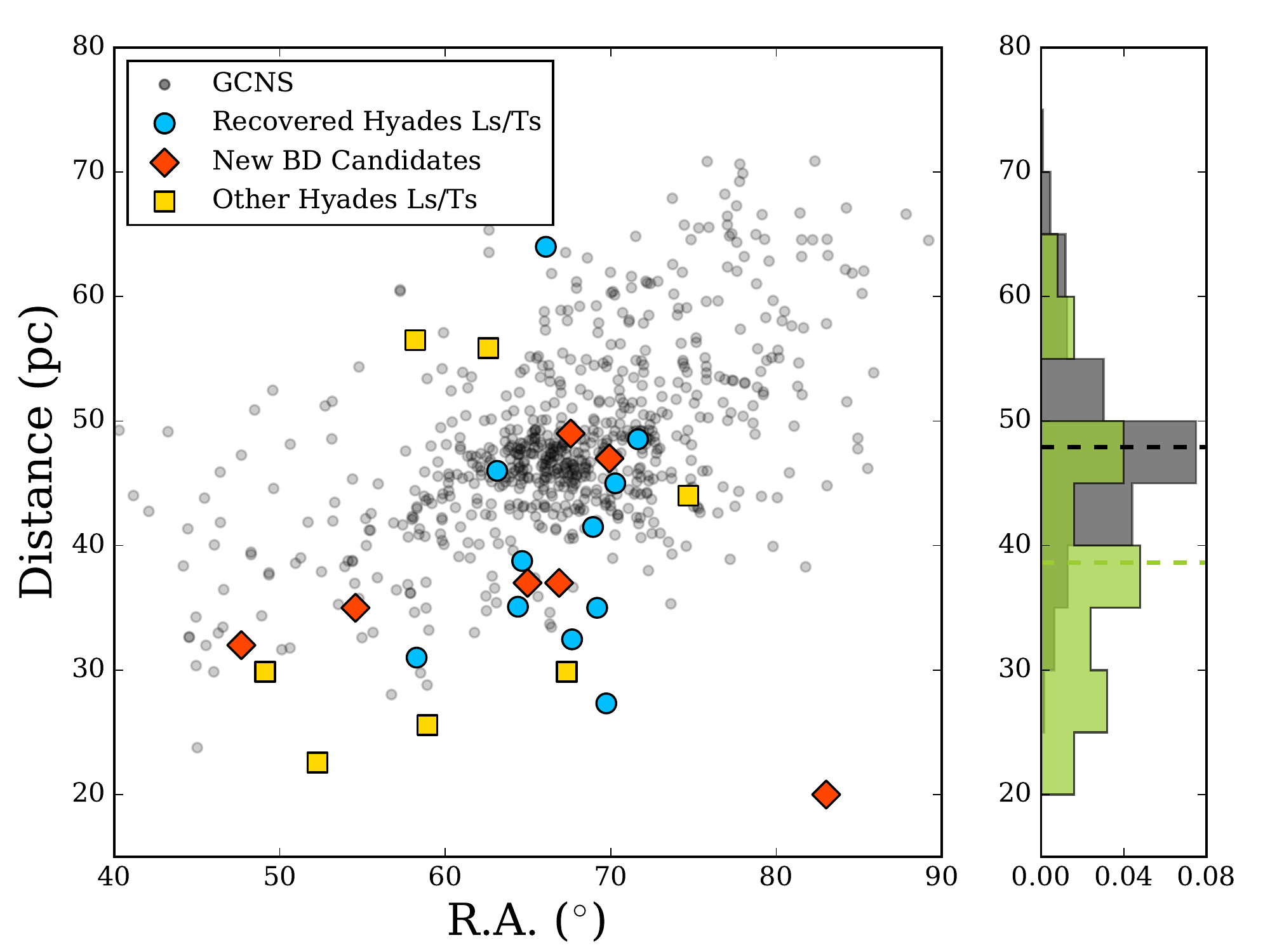}
\caption{Right ascension versus distance for the new substellar Hyades candidates in this work (red diamonds), recovered substellar Hyades members from our search (blue circles), and other suggested Hyades L and T members (yellow squares) compared to the 713 Hyades members within a 30 pc radius from the cluster center from the GCNS \citep{smart2021}.  The normalized histograms on the right show the 713 Hyades members from \cite{smart2021} in grey, and the combined sample of all substellar candidates in green.  The median values of each sample are marked with dashed lines.  }  
\label{fig:mass}
\end{figure*}

Figure \ref{fig:mass} shows the full substellar candidate sample (25 total) versus the 713 member census from \cite{smart2021}.   While substellar candidates have been found throughout the cluster, many occupy its nearest edge.  As shown in the histograms on the right side of Figure \ref{fig:mass}, the median distances of the \cite{smart2021} sample and the combined substellar sample peak at different values.  The average distance to members from the \cite{smart2021} sample is 47.9 pc, while the average distance to the substellar Hyades candidates sample is 38.6 pc.  If we exclude object's with low membership probabilities ($<$50\%) from BANYAN~$\Sigma$ (CWISE J031042.59$+$204629.3 (47.6\%), Hya02 (23.9\%), CWISE J035304.34$+$041820.0 (14.7\%), and CWISE J053204.60$+$111955 (7.1\%)), the average distance becomes 39.4 pc.  Either way, the average distance difference between substellar candidates and known cluster members indicates that either the substellar sample is contaminated by nearby, unrelated interlopers, the full substellar population of the Hyades has yet to be explored, or some combination of the two.  Considering that the search detailed in this work implemented a $J-$mag cut of 17.5 mag, which does not allow for the full extent of spectral types to be probed throughout the entire cluster radius, a deeper search would likely return more substellar Hyades members.   

\begin{longrotatetable} 
\begin{deluxetable*}{lccccrrcccccc}
\label{tab:all}
\tablecaption{Substellar Hyades Members and Candidate Members}
\tablehead{
\colhead{CWISE} & \colhead{Other} & \colhead{Disc.} & \colhead{SpT} & \colhead{SpT} & \colhead{$\mu_{\alpha}$} & \colhead{$\mu_{\delta}$} & \colhead{$\mu$} & \colhead{Dist.\tablenotemark{a}} & \colhead{Dist.} & \colhead{Mass} & \colhead{BANYAN}\\
\colhead{Name} & \colhead{Name} & \colhead{Ref.} & \colhead{} & \colhead{Ref.} & \colhead{(mas yr$^{-1}$)} & \colhead{(mas yr$^{-1}$)} & \colhead{Ref.} & \colhead{(pc)} & \colhead{Ref.} & \colhead{($M_{\rm Jup}$)} & \colhead{(\%)}  }
\startdata
J031042.59$+$204629.3 & \dots & 1 & L5 & 1 & 164.7$\pm$2.2 & -27.7$\pm$2.2 & 1 & [32] & 1 & 49$\pm$6 & 47.7\\
J031627.87$+$265027.2 & PSO J049.1159$+$26.8409 & 2 & T2.5 & 2 & 201.1$\pm$2.4 & -52.8$\pm$1.9 & 3 & 29.9$\pm$2.8 & 3  & 33$\pm$3 & 85.6 \\
J032905.95$+$132231.5 & PSO J052.2746$+$13.3754 & 3 & T3.5 & 3 & 273.2$\pm$2.0 & -20.7$\pm$2.0 & 3 & 22.6$\pm$1.5 & 3  & 33$\pm$3 & 92.5\\
J033817.87$+$171744.1 & \dots & 1 & L7 & 1 & 152.0$\pm$3.5 & -25.3$\pm$3.3 & 1 & [35] & 1 & 41$^{+6}_{-5}$ & 83.6\\
J035246.42$+$211232.7 & Hya02 & 4 & L1.5 & 5 & 116.4$\pm$2.0 & -26.9$\pm$1.5 & 6 & 56.5$\pm$6.4 & 6 & 65$\pm$6  & 24.4 \\ 
J035304.34$+$041820.0 & 2MASS J03530419$+$0418193 & 7 & L6pec (red) & 7 & 171.2$\pm$3.1 & 35.8$\pm$2.9 & 1 & [31] & 1 & 45$\pm$6 & 14.7  \\
J035542.11$+$225700.9 & Hya11 & 4 & L3 & 8 & 164.4$\pm$3.0 & -41.1$\pm$2.3 & 9 & 25.6$\pm$10.7 & 6 & 58$\pm$6 & 82.0 \\ 
J041024.02$+$145910.1 & Hya03 & 4 & L0.5 & 5 & 110.2$\pm$1.4 & -13.1$\pm$1.3 & 10 & 55.9$\pm$3.7 & 10 & 70$\pm$6 & 72.6\\
J041232.79$+$104408.0 & WISEA J041232.77$+$104408.3 & 11 & L5:~(red) & 11 & 129.5$\pm$3.8 & -5.5$\pm$3.5 & 1 & [46] & 11 & 49$\pm$6 & 96.2 \\
J041733.97$+$143015.2 & Hya10 & 4 & L2\tablenotemark{b} & 6 & 123.6$\pm$2.7 & -17.8$\pm$2.3 & 1 & 35.1$\pm$4.8 & 6  & 62$\pm$6 & 99.2  \\
J041835.00$+$213126.6 & 2MASS J04183483$+$2131275 & 12 & L5 & 12 & 141.5$\pm$2.7 & -45.7$\pm$2.3 & 6 & 38.8$\pm$4.4 & 6 & 49$\pm$6 & 98.3 \\
J041953.55$+$203628.0 & \dots & 1 & T4 & 1 & 109.4$\pm$9.0 & -35.8$\pm$8.9 & 1 & [37] & 1 & 32$\pm$3 & 93.3 \\
J042418.72$+$063745.5 & 2M0424$+$0637 & 13 & L4 & 13 & 138.8$\pm$2.8 & 7.5$\pm$2.9 & 1 & [64] & 1 & 53$\pm$6 & 93.4 \\
J042731.38$+$074344.9 & \dots & 1 & L7 & 1 & 114.3$\pm$3.5 &  5.5$\pm$3.1 & 1 & [37] & 1 & 41$^{+6}_{-5}$ & 90.0 \\
J042922.88$+$153529.4 & CFHT-Hy-21 & 14 & T1 & 14 & 82.1$\pm$9.8 & -15.5$\pm$8.6 & 6 & 29.9$\pm$11.3 & 6 & 34$\pm$3 & 74.7 \\  
J043018.70$+$105857.1 & \dots & 1 & T3 & 1 & 106.3$\pm$6.9 & -10.7$\pm$6.9 & 1 & [49] & 1 & 33$\pm$3 & 94.4 \\
J043038.87$+$130956.7 & CFHT-Hy-20 & 14 & T2.5 & 15 & 142.6$\pm$1.6 & -16.5$\pm$1.7 & 15 & 32.5$\pm$1.6 & 15 & 33$\pm$3 & 98.7 \\
J043543.04$+$132344.8 & Hya12 & 4 & L6 (red)\tablenotemark{c} & 2 & 100.2$\pm$1.9 & -15.1$\pm$2.0 & 6 & 41.5$\pm$3.6 & 6 & 45$\pm$6 & 99.5 \\
J043642.79$+$190134.6 & WISEA J043642.75$+$190134.8 & 11 & L6 & 11 & 113.5$\pm$2.0 & -42.1$\pm$2.0 & 1 & [35] & 11 & 45$\pm$6 & 96.2 \\
J043855.29$+$042300.6 & PSO J069.7303$+$04.3834 & 3 & T2 & 3 & 118.7$\pm$3.5 & 11.7$\pm$3.4 & 1 & 27.3$\pm$4.3 & 3 & 34$\pm$3 & 86.0 \\
J043941.41$+$202514.8 & \dots & 1 & T3 & 1 & 80.8$\pm$8.0 & -30.3$\pm$7.9 & 1 & [47] & 1 & 33$\pm$3 & 62.0 \\
J044105.60$+$213001.3 & WISEA J044105.56$+$213001.5 & 11 & L5 (red) & 11 & 98.7$\pm$4.6 & -48.5$\pm$4.4 & 1 & [45] & 11 & 49$\pm$6 & 90.5\\
J044635.44$+$145125.7 & Hya19 & 4 & L3.5 & 5 & 79.1$\pm$2.6 & -22.4$\pm$2.5 & 1 & 48.5$\pm$5.9 & 6 & 55$\pm$6 & 98.5 \\
J045845.76$+$121234.1 & Hya08 & 4 & L0.5 & 5 & 88.6$\pm$1.1 & -17.5$\pm$0.8 & 10 & 44.0$\pm$1.7 & 10 & 70$\pm$6 & 99.0  \\
J053204.60$+$111955.1 & \dots & 1 & L7 & 1 & 72.4$\pm$1.9 & -30.2$\pm$1.9 & 1 & [20] & 1 & 41$^{+6}_{-5}$ & 7.1 \\
\enddata
\tablenotetext{a}{Distances in square brackets are photometric distances.  All other distances come from measured parallaxes.}
\tablenotetext{b}{\cite{lodieu2014} report an optical spectral type of L1 for this object, while \cite{martin2018} found an optical spectral type of L3.5.  \cite{lodieu2019} gives a spectral type of L2, and we adopt that type here.}
\tablenotetext{c}{\cite{lodieu2014} found an optical spectral type of L3.5 for this object.}
\tablerefs{(1) This work; (2) \cite{best2015}; (3) \cite{best2020}; (4) \cite{hogan2008}; (5) \cite{lodieu2014}; (6) \cite{lodieu2019}; (7) \cite{kellogg2017}; (8) \cite{martin2018}; (9) \cite{best2018}; (10) \cite{gaia2021}; (11) \cite{schneider2017}; (12) \cite{perez2017}; (13) \cite{perez2018}; (14) \cite{bouvier2008}; (15) \cite{liu2016}}
\end{deluxetable*}
\end{longrotatetable} 

\section{Summary}
We have presented a search for substellar members of the Hyades based on data from the UKIRT Hemisphere Survey.  We found 25 candidates, 10 of which were previously suggested substellar Hyades members.  We classified two known brown dwarfs recovered in our search as potential Hyades members.  Of the 13 new discoveries, 6 objects were found to be unrelated, background cool stars.  Five new discoveries are considered strong Hyades candidates, while Hyades membership cannot be ruled out for two additional discoveries.  Parallax and radial velocity measurements will be necessary to confirm Hyades membership for all of these candidates.  We also find that the current census of substellar Hyades candidates is likely incomplete, and deeper searches, using the UHS or other surveys, would reveal a more complete accounting of the substellar Hyades population.

\acknowledgments
This publication makes use of data products from the UKIRT Hemisphere Survey, which is a joint project of the United States Naval Observatory, The University of Hawaii Institute for Astronomy, the Cambridge University Cambridge Astronomy Survey Unit, and the University of Edinburgh Wide-Field Astronomy Unit (WFAU).  This project was primarily funded by the United States Navy.  The WFAU gratefully acknowledges support for this work from the Science and Technology Facilities Council through ST/T002956/1 and previous grants.  The authors acknowledge the support provided by the US Naval Observatory in the areas of celestial and reference frame research, including the USNO's postdoctoral program.  This material is supported by the National Science Foundation under Grant No. 2007068, 2009136, and 2009177.  This publication makes use of data products from the {\it Wide-field Infrared Survey Explorer}, which is a joint project of the University of California, Los Angeles, and the Jet Propulsion Laboratory/California Institute of Technology, and NEOWISE which is a project of the Jet Propulsion Laboratory/California Institute of Technology. {\it WISE} and NEOWISE are funded by the National Aeronautics and Space Administration.  Part of this research was carried out at the Jet Propulsion Laboratory, California Institute of Technology, under a contract with the National Aeronautics and Space Administration.  The authors wish to recognize and acknowledge the very significant cultural role and reverence that the summit of Maunakea has always had within the indigenous Hawaiian community.  We are most fortunate to have the opportunity to conduct observations from this mountain.

\facilities{UKIRT, IRTF}

\software{BANYAN~$\Sigma$ \citep{gagne2018}, SpeXTool \citep{cushing2004}}

\end{document}